\documentclass[prd,twocolumn,amssymb,amsmath]{revtex4}

\usepackage{graphicx}
\usepackage{txfonts}

\usepackage{color}

\begin{document}
\newcommand{\bx}{{\bf x}}
\newcommand{\bn}{{\bf n}}
\newcommand{\bk}{{\bf k}}
\newcommand{\dd}{{\rm d}}
\newcommand{\dslash}{D\!\!\!\!/}
\def\ga{\mathrel{\raise.3ex\hbox{$>$\kern-.75em\lower1ex\hbox{$\sim$}}}}
\def\la{\mathrel{\raise.3ex\hbox{$<$\kern-.75em\lower1ex\hbox{$\sim$}}}}
\def\beq{\begin{equation}}
\def\eeq{\end{equation}}

\vskip-2cm
\title{\textcolor{black}{Axions and Atomic Clocks}}

\author{Lawrence M. Krauss}
\affiliation{
   The Origins Project Foundation, Phoenix AZ 85064}
\email{krauss@asu.edu}

\vspace*{1cm}

\begin{abstract} 

The equations of electrodynamics are altered in the presence of a classical coherent axion dark matter background field, changing the dispersion relation for electromagnetic waves.  Careful measurements of the frequency stability in sensitive atomic clocks could in principle provide evidence for such a background for $f_a \ge 10^7$ GeV.  Turning on a background magnetic field might enhance these effects in a controllable way, and interferometric measurements might also be useful for probing the time-varying photon dispersion relation that results from a coherent cosmic axion background.
\end{abstract}

\date{\today}

\maketitle



\section{Introduction} 
\label{introduction}

 Axions, postulated to exist as a consequence of the proposed Peccei-Quinn solution of the strong CP problem \cite{pq}, are also a favored candidate for the dark matter that dominates the matter density of the Universe \cite{dm}.  Unlike most other cold dark matter candidates, an axion cold dark matter background would be comprised of a vacuum condensate rather than a thermal Boltzmann distribution of individual particles.  As long as coherence is maintained, and momenta are small, the axion background can therefore be treated as a classical background field 
 
Because of their connection to anomalies and CP violation, axions are coupled to the electromagnetic field by the coupling

\begin{equation}
\frac{g}{4}a F^{\mu\nu}\tilde{F}_{\mu\nu},
\label{eq:111}
\end{equation}
where units are chosen so that  $g_a$ is dimensionless.  In this case, Maxwell's equations are modified in the presence of a background axion field, $a=a_0\sin\omega_a t$, which is normalized so that its contribution to the global energy density today is $m_a^2a_0^2/2=\rho_{\text{DM}}=3.6\times10^{-42}{GeV^4}$  ~\cite{PDG}, to be,

\begin{equation}
\vec{\nabla}\cdot\vec{E} = \rho - g \vec{B}\cdot\vec{\nabla}a 
\end{equation}
\begin{equation}
\vec{\nabla}\times\vec{B} -\frac{\partial\vec{E}}{\partial t}= \vec{j} - {g}\left(\frac{\partial a\vec{B}}{\partial t} +\vec{\nabla} \times a\vec{E} \right),
\label{eq:2}
\end{equation}
where $\vec{j}$ is the electric current density~\cite{sikiviewilczek}. In vacua, if spatial gradient of $a$ are small on the scale of an experimental apparatus, which is likely, assuming small axion momenta, then the terms involving $\vec{\nabla}a$ can be ignored so that, using Maxwell's equation for the curl of $E$, which is unaffected by the axion field  (\ref{eq:2}) becomes

\begin{equation}
\vec{\nabla}\times\vec{B} -\frac{\partial\vec{E}}{\partial t}=  -{g}\frac{\partial a}{\partial t}\vec{B} 
\label{eq:4} ,
\end{equation}

 . 


\section{Electromagnetic Waves in a background axion field} 
\label{photons}

The presence of the non-zero term term on the RHS of (\ref{eq:4}) has two effects on the propagation of electromagnetic waves.  First, if $\vec{B}$ represents the oscillating magnetic field of an electromagnetic wave and if the RHS of the equation is zero, as it is in standard electrodynamics without the presence of external sources, linear polarized waves will propagate unaffected.  However, note that for such a linearly polarized wave addition of the term on the RHS of  (\ref{eq:4}) changes the dynamics. In this case, this term is perpendicular to the two terms on the LHS, so that the presence of an axion background serves to rotate the polarization of EM waves.   

In the presence of such a background the eigenmodes of electromagnetic waves will be circularly polarized, with the sign of the term on the RHS of  (\ref{eq:4}) switching for left vs right circularly polarized waves. 

The presence of a non-zero term on the RHS of  (\ref{eq:4}) has another effect, which can most easily be seen if we use the circular polarization basis mentioned above.   This term will change the dispersion relation for EM waves (see \cite{renau} for a related analysis), so that if we take a spatial Fourier transform and expand in momentum modes, the standard relation ${w}^2 ={k}^2$ in the absence of an axion background becomes, treating the background effect as a small perturbation 

\begin{equation}
{\omega}^2 (1 \pm {gm_a{a(t)}\over \omega} )={k}^2 
\label{eq:3}  
\end{equation}
where the two signs apply to the two different circular polarization modes. A coherent axion background therefore introduces a time-dependent dispersion relation for electromagnetic waves--effectively acting like a time-dependent fluctuating mass for the photon, with a time-averaged value of zero, but with non-zero variance.

To get a sense of the magnitude of this effect, using the relation 
\begin{equation}
g = 2g_{a\gamma\gamma}\alpha/{\pi} \times  [1/f_a]
\end{equation}
for $g_{a\gamma\gamma} \approx 1$, a range $f_a \approx 10^7 -10^{14}$ GeV this implies  $gm_aa_0 \approx 10^{-21} - 10^{-28}$ eV.

This effect is well below direct and astrophysical probes for a photon mass, which currently give upper limits in the range of $10^{-14}-10^{-18} $eV \cite{photons}, although it has been suggested as a possible mechanism affecting high energy cosmic rays \cite{andrianov}

\section{ Axions and frequency fluctuations in Atomic Clocks}

Atomic Clocks, based on ultra-precision quantum optics technology, have reached long term timing precisions on the order of one part in $10^{19}$ \cite{jila} and are continuing to improve.  This allows unprecedented sensitivity to ultra-weak noise backgrounds that may impact on frequency stability.   This stability is frequency-dependent however.  The original atomic clocks involved microwave frequency resonant cavities and masers, and these have achieved long time frequency stability on the order of one part in $10^{16}$ \cite{atommicrowave}.  Atomic physics techniques involving laser cooling of trapped atoms and ions has led to dramatic improvements in precision involving optical frequency clocks, and although measurements are harder with these systems, the use of frequency combs and other techniques has successfully addressed these challenges \cite{jila}. 

The extremely narrow line-widths necessary to achieve such timekeeping accuracy suggests the possibility, in principle, of probing for a cosmic axion background through its impact on the photon dispersion relation described above 

From (\ref{eq:3}) the lower the frequency of electromagnetic radiation the larger the relative impact of an oscillating background axion field is on the frequency-momentum phase relation.   The figure of merit is 

\begin{equation}
\kappa = {gm_a{a(t)}\over \omega}.   
\end{equation}

For microwaves, with energies of order $10^{-5}$ eV, $\kappa \approx 10^{-16}$ for  $f \approx 10^7$ GeV, so that traditional atomic clocks might already, at least in principle, be sensitive to the high end range of interesting cosmological axion masses.  For optical frequencies, or order $1$ eV, $\kappa \approx 10^{-21}$ for $f \approx 10^7$ GeV, which is about 2 orders of magnitude smaller than the existing precision of optical atomic clocks might probe, but the sensitivity of these clocks is improving rapidly. .

A time-varying photon dispersion relation would result in either line broadening, or the appearance of side-lobes, both of which are tightly controlled in atomic clocks.   Whether or not it would be is possible in practice for atomic clocks to specifically search for such a signal due to axion masses of cosmological interest will depend on detailed experimental configurations.  

The most important variable that will be relevant to experimental design is likely to be the ratio of the time variation of the axion field to the frequency of the electromagnetic wave of use in the clock.  For example, for $f \approx 10^7$ GeV, $m_a \approx  0.1$ eV, so that the frequency of variation of the axion field is approximately $10^5$ times larger than the frequency of electromagnetic radiation in microwave based atomic clocks.  For optically based atomic clocks the associated axion variational frequency is smaller for almost all the entire interesting axion mass range.   Thus, for microwaves, propagating electromagnetic wave will experience a fluctuating axion field over a single wavelength, and will thus be sensitive to variance in that field.  A $\sqrt n$ reduction in magnitude for the effective variation in the propagating EM dispersion relation would put predicted axion background effects beyond the range of current microwave atomic clock sensitivities, but perhaps not for experiments specifically designed to search for this variation.   For optical wavelengths, waves will experience a constant or slowly varying axion field over single cycle, but if they propagate over many cycles, the time varying dispersion relation should have a direct effect on the atomic clock precisions. 

The fact that the axion background produces a time varying photon dispersion relation also suggests the possibility that interferometric techniques might be useful in this regard, especially if the photon beams traverse distances that may be long compared to the axion coherence length.  Since the axion bandwidth is roughly $10^{-6}$ due to the small relative momentum of the axion field arising from its motion in the galaxy (with $v^2/c^2 < 10^{-6}$), this coherence length will depend on the axion frequency being probed. 

More detailed estimates of possible sensitivity of actual atomic clock-based direct or interferometric axion search experiments to a cosmological axion background will depend specific experimental configuration issues that are beyond the scope of this letter.  The estimates presented here suggest that rapidly improving atomic clock precisions could in principle allow sensitivity to a cosmic axion background now or in the near future, and also motivate considerations of adapting current experiments for this purpose. 

\section{ Axions and electromagnetic propagation in the presence of external magnetic fields}

Detection of a coherent cosmic axion background directly using atomic clock sensitivities is challenging, being at or just beyond the limits of current sensitivities.  We have also not considered directly the challenges of isolating this signal from other possible sources of noise.  The fact that the variation in the photon dispersion relation is at a fixed frequency with narrow bandwidth should help in isolating this effect.  Of course, extracting a signal from other noise would be even easier if one could somehow turn the effect of the cosmic axion background on and off.  In empty space this is not possible.  However incorporating an additional fixed background magnetic field could might allow one to impact the propagation of electromagnetic waves in atomic clocks or other resonant cavities..  From (\ref{eq:4}) it is clear that an axion background not only affects the dispersion relation of EM waves but it also allows the possibility of mixing photons of different frequencies.  First consider a sufficiently intense additional static background magnetic field $\vec{B_0}$ field.  If $\vec{B_0} >> \vec{B}$ then \ref{eq:4} becomes

\begin{equation}
\vec{\nabla}\times\vec{B} -\frac{\partial\vec{E}}{\partial t} \approx  -{g}\frac{\partial a}{\partial t}\vec{B_0} \approx j_{eff} 
\label{eq:5} ,
\end{equation}
where $j_{eff} = gm_aa(t)B_0$.  Namely, an axion background couples to the magnetic field to act as an effective oscillating current density  

This effect has currently been proposed as a method to detect axions, via this effective oscillating current using sensitive SQUID technology \cite{sushkov}.  However it is worth noting that this oscillating current will also be the source of oscillating electromagnetic fields.  Could such appropriately prepared fields interfere in a measurable way with the finely tuned electromagnetic waves used in atomic clocks if the axion frequency is close to the laser frequency?

Another possibility, although more challenging from an experimental point of view, is to consider an initial time-varying external magnetic field in a cavity traversed by a laser beam. In this case, taking the curl of both sides of (\ref{eq:4}), and again assuming the external field is much larger than the field due to the incident beam,

\begin{equation}
\nabla^2\vec{B} + \frac{\partial^2\vec{B}}{\partial t^2} \approx -g \frac {\partial a}{\partial t} \frac{\partial \vec{E_0}}{\partial t}
\end{equation}
where $\vec{E_0}$ is the effective electric field produced by the time varying $\vec{B_0}(t)$ field in the cavity.  By varying the frequency of the external magnetic field variation, one might match the product of the axion frequency and the external field variation frequency with the square of the incident electromagnetic wave frequency in a way to once again effectively produce a a time varying dispersion relation for the incident electromagnetic wave, but in this case one in which the effect of the axion coherent background could be enhanced by the something like the ratio $B_0/B$.   Physically this effect could be considered as electromagnetic waves associated with the background oscillating field in the cavity scattering off the background axion field and down-converting into photons that interfere with the incident laser beam.


\section{Conclusions and outlook} 
\label{conclusions}

Direct axion-photon conversion has thus far been the chief method explored for probing for the existence of a cosmic axion dark matter background (\i.e. \cite{sikivielmk}).  As we have stressed here, however, this background also has an effect on photon propagation in the vacuum that might be measurable.  The axion background produces an effective time-varying dispersion relation for electromagnetic waves--acting to some extent like a time-varying photon mass term.  While this effect appears to be too small, by at least 3 orders of magnitude to be detectable for existing studies of photons from known astrophysical sources, the effect appears to be in the range of the sensitivity of current atomic clocks for a Peccei-Quinn symmetry breaking scales that may be relevant for a cosmological axion background, and within a few orders of magnitude of that for optical clocks which could probe for higher symmetry breaking (and lower axion mass) scales.  To probe for these effects, experiments would have to be designed to probe for specific narrow frequency  noise in these detectors, with different requirements when the axion frequency is greater or less than the incident electromagnetic wave frequency.  The effect of actions on photon dispersion relations might be enhanced by introducing additional static or time varying external magnetic fields into the system, but it is not clear whether this would be experimentally feasible in sensitive atomic clocks.  Finally, since an axion background effectively changes the dispersion relation for electromagnetic waves, it is worth considering whether sensitive interferometers might be adapted for the purpose of detecting this effect for cosmic axions.


\section*{Acknowledgments}

{LMK acknowledges the hospitality of the IEEE International Frequency Control Symposium and the European Frequency and Time Forum in Besan\c{c}on France in 2017 where he first learned about the detailed potential of atomic clocks.  He also thanks Alex Sushkov for useful conversations about an initial draft of this letter. }



\begin{thebibliography}{99}























\bibitem{pq}
Peccei, R. D. and Quinn, H. R., Phys. Rev. Lett. 38, 1440  (1977), S. Weinberg, Phys. Rev. Lett. 40, 223 (1978), F. Wilczek, Phys. Rev. Lett. 40, 279 (1978).

\bibitem{dm}
Abbott, L. and Sikivie, P., Phys. Lett. B120, 133 (1983), Preskill, J., Wise, M.B., Wilczek, F., Phys. Lett B120, 127 (1983), Dine, M., Fischler, W.. Phys.Lett. B120 (1983) 137.

\bibitem{PDG}
Tanabashi, M. {\it et al.} (Particle Data Group), Phys. Rev. D 98, 030001 (2018).

\bibitem{sikiviewilczek}
Sikivie, P., Phys. Rev. Lett. 51, 1415 (1983), Wilczek, F., Phys. Rev. Lett. 58, 1799 (1987).


\bibitem{renau}
Espriu, D and Renau, A., Phys. Rev. D85, 025010 (2012), Renau, A., arXiv  e print 1512.03311 (2015) .

\bibitem{photons}
Bonetti, L. {\it et al}, Phys. Lett B757, 548 (2016), Olive, K.A. et al., Particle Data Group Collaboration
Review of particle physics,
Chin. Phys. C, 38 , 090001, (2014).

\bibitem{andrianov}
Andrianov, A. A. and Espriu, D. and Mescia, F. and Renau, A., Phys. Lett. B684 101 (2010).

\bibitem{jila}
Marti G.E. {\it et al}, Phys. Rev. Lett 120 (10): 103201 (2018). 


\bibitem{atommicrowave}
BIPM Annual Report on Time Activities, Volume 10  ISBN 978-92-822-2263-8  (2015).

\bibitem{sushkov}
Gramolin, A.V. {\it et al}, arXiv e print  1811.03231 (2018).


\bibitem{sikivielmk} 
P. Sikivie Phys. Rev. Lett. 52, 695 (1984),
Krauss, L.M., Moody, J., Wilczek, F., Morris, D. E. Phys. Rev. Lett 55(17), 1797-1800 (1985).



     \end{thebibliography}
\end{document}